\documentclass[twocolumn]{aastex7}
\usepackage{amsmath}
\usepackage{amssymb}
\usepackage{apjfonts}

\newlength{\apjcolwidth}
\newcommand{\afe}{{\rm[\alpha/Fe]}}
\newcommand{\feh}{{\rm[Fe/H]}}

\begin{document}

\title{MESA Isochrones and Stellar Tracks (MIST) III. The White Dwarf Cooling Sequence}

\author[0000-0002-4791-6724]{Evan B. Bauer}
\affiliation{Lawrence Livermore National Laboratory, Livermore, California 94550, USA}
\affiliation{Center for Astrophysics $|$ Harvard \& Smithsonian, 60 Garden Street, Cambridge, MA 02138, USA}
\email{bauer39@llnl.gov}

\author[0000-0002-4442-5700]{Aaron Dotter}
\affiliation{Department of Physics and Astronomy, Dartmouth College, 6127 Wilder Laboratory, Hanover, NH 03755, USA}
\email{Aaron.L.Dotter@dartmouth.edu}

\author[0000-0002-1590-8551]{Charlie Conroy}
\affiliation{Center for Astrophysics $|$ Harvard \& Smithsonian, 60 Garden Street, Cambridge, MA 02138, USA}
\email{cconroy@cfa.harvard.edu}

\author[0000-0001-7296-3533]{Tim Cunningham}
\altaffiliation{NASA Hubble Fellow}
\affiliation{Center for Astrophysics $|$ Harvard \& Smithsonian, 60 Garden Street, Cambridge, MA 02138, USA}
\email{tim.cunningham@cfa.harvard.edu}

\author[0000-0002-8435-9402]{Minjung Park}
\affiliation{Center for Astrophysics $|$ Harvard \& Smithsonian, 60 Garden Street, Cambridge, MA 02138, USA}
\email{minjung.park@cfa.harvard.edu}

\author[0000-0001-9873-0121]{Pier-Emmanuel Tremblay}
\affiliation{Department of Physics, University of Warwick, Coventry, CV4 7AL, UK}
\email{P.Tremblay@warwick.ac.uk}

\begin{abstract}
  We present a substantial update to the MESA Isochrones and Stellar
  Tracks (MIST) library,
  extending the MIST model grids and isochrones down the WD
  cooling sequence with realistic physics for WD cooling timescales.
  This work provides a large grid of MESA models for Carbon-Oxygen
  core WDs with hydrogen atmospheres (spectral type DA/DC),
  descended from full prior stellar evolution calculations.
  The model tracks, isochrones, and WD cooling timescale contours are
  available on the MIST project website
  and at\dataset[10.5281/zenodo.15242047]{\doi{10.5281/zenodo.15242047}}.
  Our WD models provide a very large, publicly available grid
  with detailed physics for WD cooling timescales: realistic interior
  and envelope compositions, with element diffusion and heavy-element
  sedimentation, nuclear burning at the base of the WD hydrogen
  envelope, core crystallization, and C/O phase separation.
  As a large grid of open-source stellar evolution models,
  these WD models provide both out-of-the-box model tracks for
  comparison with observations and a framework for building further WD
  models to investigate variations in WD physics.
\end{abstract}

\keywords{Stellar physics (1621), White dwarf stars (1799)}

\section{Introduction}

Over one hundred years ago, 40 Eridani B was the first white dwarf star
to appear below the main sequence on a Hertzsprung--Russell (HR) diagram
\citep{Russell1914,Hertzsprung1915}. We now have detailed measurements of
magnitudes, colors, and distances for hundreds of thousands of white
dwarfs (WDs) within a few hundred pc \citep{GaiaHRD,GentileFusillo2021}.
The emerging structure of white dwarf cooling sequences in the data is
revealing a great deal about stellar physics
\citep{Tremblay2019,Tremblay2024,Cheng2019,Saumon2022,Heintz2022,Barrientos2024,Bedard2024,Pathak2025},
star formation histories \citep{Winget1987,Roberts2025}, stellar populations
\citep{vonHippel2005,Hansen2007,Kalirai2007,Bedin2008,Bedin2019,Bedin2025,Moss2022,Salaris2024},
and more.

\begin{figure*}
  \centering
  \includegraphics{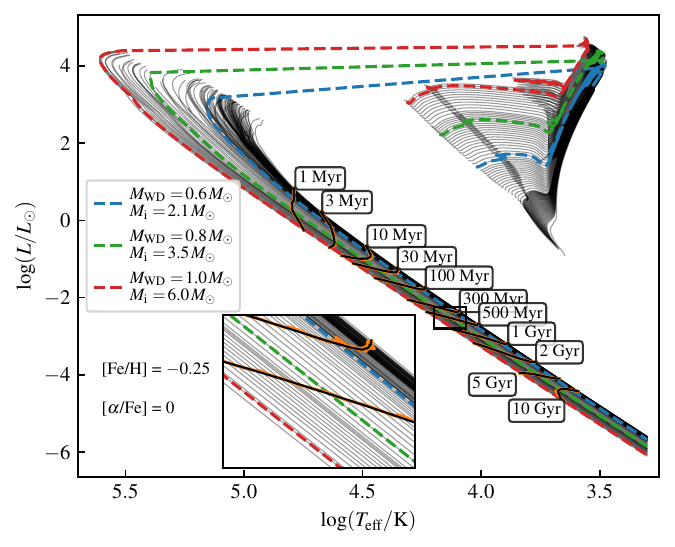}
  \includegraphics[width=\apjcolwidth]{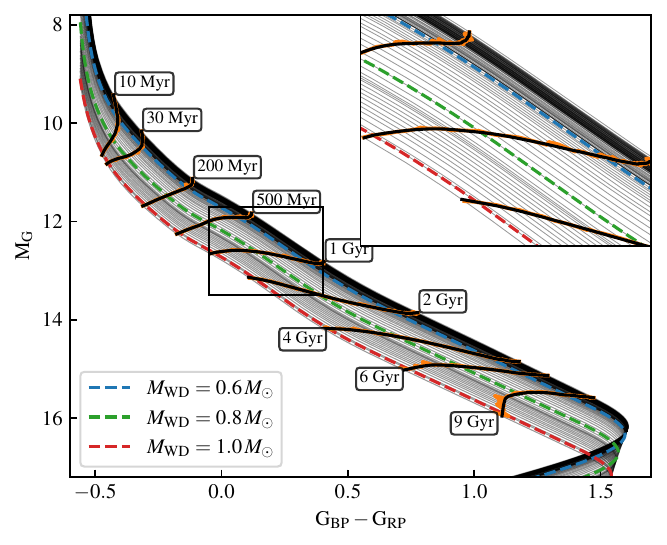}
  \includegraphics[width=\apjcolwidth]{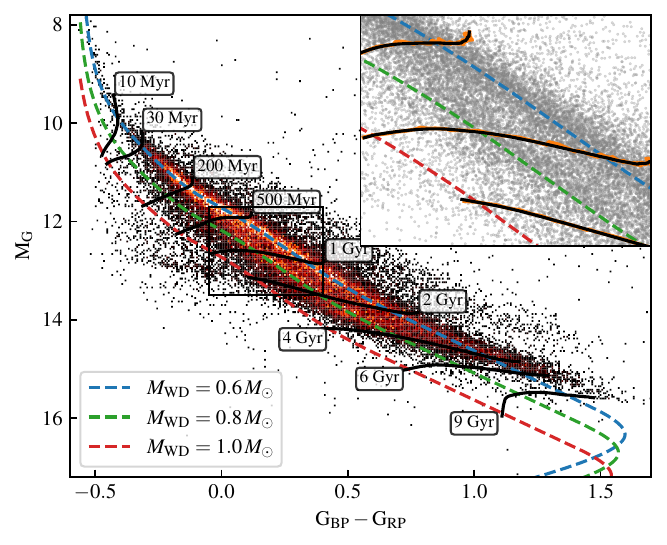}
  \caption{HR diagram (top) and {\it Gaia} CMDs (bottom) showing
    evolutionary tracks and WD cooling age contours for the 100 WD
    models descended from $\feh = -0.25$, $\afe = 0$ progenitors.
      The lower two panels show parts of the same model grid at
      this metallicity on the {\it Gaia} CMD, separated for clarity to
      emphasize the density of the full model grid in the left panel,
      while comparing just a few representative tracks to the
      {\it Gaia} WD data in the right panel.
      Black cooling contour curves are smoothed (see text), while
      the underlying non-smoothed contours are displayed in orange for
      comparison to show how much smoothing has taken place.
    Note that here the age contours are for WD {\it cooling} age,
    rather than {\it total} stellar age.
    Example contours for constant total age are shown in the
    isochrones of Figure~\ref{fig:isochrones}.
  }
  \label{fig:HRtracks_all}
\end{figure*}

Having exhausted nuclear fuel as a luminosity source, WDs radiate away
their thermal energy at predictable rates \citep{Mestel1952}, making
them useful as ``cosmic chronometers''
\citep{Winget1987,Fontaine2001}, either at the stellar population
level or in binaries with stellar companions. Grids of computational
models for WD cooling are widely used in the literature for this
purpose (e.g., those of \citealt{Renedo2010,Bedard2020,Salaris2022}).
The stellar evolution code Modules for Experiments in Stellar Astrophysics (MESA,
\citealt{Paxton2011,Paxton2013,Paxton2015,Paxton2018,Paxton2019,Jermyn2023})
is capable of evolving WD models with state of the art physics
\citep{Bauer2019,Bauer2020,Bauer2023}, but so far no comprehensive
grid of WD evolution models using MESA has been widely available. In
this work, we provide such a model grid within the framework of the
MESA Isochrones and Stellar Tracks (MIST) project. MIST was first
introduced in \cite{Dotter2016} (Paper~0) and \cite{Choi2016}
(Paper~I), providing a large grid of MESA stellar models and isochrones evolving from
the pre-main sequence over the mass range $0.1-300\,M_\odot$ and
metallicity range $-4 \leq \feh \leq 0.5$. MIST also provides
synthetic photometry options for tracks and isochrones in
addition to the theoretical output of the MESA models. Recently,
Dotter et al.~(submitted, Paper~II) have released a new revision of the
MIST model grids with substantial physics updates and an
additional metallicity dimension $-0.2 \leq \afe \leq 0.6$ for better
describing stellar populations. Here we extend that work by continuing
the evolution of those models through the white dwarf cooling sequence
(WDCS) to provide WD cooling timescales and include this evolutionary
stage in the MIST isochrones.

We begin in Section~\ref{s.coverage} with a broad overview of the
thousands of WD model tracks included in this work, which are
descended from the full progenitor evolution models of
Paper~II. Section~\ref{s.physics} details the physics
included in our WD evolution calculations. Section~\ref{s.composition}
describes the composition of both the cores and envelopes produced by
our WD evolution calculations, and how these configurations impact the
physics of the cooling models. Section~\ref{s.comparisons} outlines
some comparisons between other WD model grids available in the
literature with the cooling timescales in our models.
We emphasize that all of the model tracks, isochrones, and WD cooling age
contours described in this work are readily available on the MIST
website
and at\dataset[10.5281/zenodo.15242047]{\doi{10.5281/zenodo.15242047}},
and that the code to reproduce all WD models and further build upon
this work using the open-source stellar evolution software MESA is
available on Zenodo:\dataset[10.5281/zenodo.15196934]{\doi{10.5281/zenodo.15196934}}.

\section{Model Coverage} \label{s.coverage}

Our white dwarf model grids are descended from the full evolutionary
sequences described in Paper~II. The model grids of Paper~II cover a
broad mass range ($0.1-300\,M_\odot$) starting from the pre-main
sequence, and span a wide range of metallicities in two dimensions:
$-4 \leq \feh \leq 0.5$ with steps of $\Delta \feh = 0.25$, and $-0.2
\leq \afe \leq 0.6$ with steps of $\Delta \afe = 0.2$.%
\footnote{
  For very low metallicity $\feh < -3$, the spacing is $\Delta\feh = 0.5$,
  so there are 17 total distinct $\feh$ values and 5 distinct $\afe$
  values in the grid. We also drop the highest metallicity at $\feh =
  +0.5$, $\afe = +0.6$ due to the lack of input physics at $Z>0.1$, so
  there are a total of $17 \times 5 - 1 = 84$ distinct compositions in
  the model grid.
}
See Papers~I and~II for a full description of the setup for these model grids
and the input physics for the stellar evolution prior to the white
dwarf phase. Low-mass models ($M<0.6\,M_\odot$) do not evolve beyond
the main sequence, and high-mass models ($M \geq 7\,M_\odot$) evolve
to core carbon burning and beyond, while our models in the mass range
$0.6\,M_\odot \leq M < 7\, M_\odot$ evolve through the AGB and
thermally-pulsing AGB phases to become white dwarf progenitors. In
Paper~II, we stopped these model tracks and isochrones at the end of
the AGB, when winds on the AGB cause the star to lose enough mass to
start evolving toward more compact radii and hotter temperatures
reaching $T_{\rm eff} = 10{,}000\,\rm K$.
In this work, we now take any model that reached this post-AGB phase
with a Carbon-Oxygen core and evolve it over to and down the WDCS.

For a given metallicity specified by $\feh$ and $\afe$, the model grid
produces $\approx 100$ C/O WD models with masses
spanning roughly $0.5 \,M_\odot < M_{\rm WD} < 1.05\,M_\odot$
descended from progenitors of mass $0.6\,M_\odot \leq M_{\rm ZAMS}
\leq 6.5\,M_\odot$.%
\footnote{Our model grids do not produce higher-mass
  ($M>1.05\,M_\odot$) Oxygen-Neon (O/Ne) WDs because our WD progenitor
  models terminate at core carbon ignition. Higher mass WD evolution
  is left for future work.}
Because our model grids include 84 ($\feh$, $\afe$) metallicity
combinations, this results in several thousand total white dwarf
evolution tracks. As an example of just one metallicity in the larger grid,
Figure~\ref{fig:HRtracks_all} shows the full HR
diagram evolution for the 100 WD models produced by our grids at $\feh
= -0.25$, $\afe = 0$, including both the progenitor evolution from
ZAMS though AGB, and then the later evolution through the cooling
sequence. We leave out the evolution across the HR diagram between the
end of the AGB and start of the WDCS because we apply some simple
unphysical settings to allow the models to quickly traverse this
short-lived but challenging phase of evolution (see
Section~\ref{s.crossing}), and this phase only lasts
$10^2$--$10^4$ years in our models.

Figure~\ref{fig:HRtracks_all} also shows these same WD evolutionary
tracks on the {\it Gaia} color-magnitude diagram (CMD), including a
panel comparing to the data for WDs within 200 pc from {\it Gaia} DR3
\citep{GaiaHRD,GaiaDR3}. By construction, these WD models are all DA/DC
type WDs with pure hydrogen atmospheres, and colors are calculated
using the synthetic spectra of \cite{Tremblay2011} (see
Section~\ref{s.BCs} for more on atmospheres and colors).
The tracks in Figure~\ref{fig:HRtracks_all} also include contours of
constant WD cooling age. Due to the density of model tracks in the
region of the WDCS, the cooling age contours can display small amounts
of noise from adjacent cooling tracks with slightly different cooling
ages, so we plot both the unsmoothed contours (orange) and contours
with LOWESS%
\footnote{
  Locally Weighted Scatterplot Smoothing \citep{Cleveland79}.
}
smoothing applied (black). We apply this smoothing to the
luminosity along the track, then recalculate $T_{\rm eff}$ from the
smoothed luminosity assuming $M$, $R$, and $\log g$ are fixed for a
given WD track. We provide data files for these smoothed cooling age
contours for all model grids in this work:%
\dataset[10.5281/zenodo.15242047]{\doi{10.5281/zenodo.15242047}}.

\begin{figure}
  \includegraphics{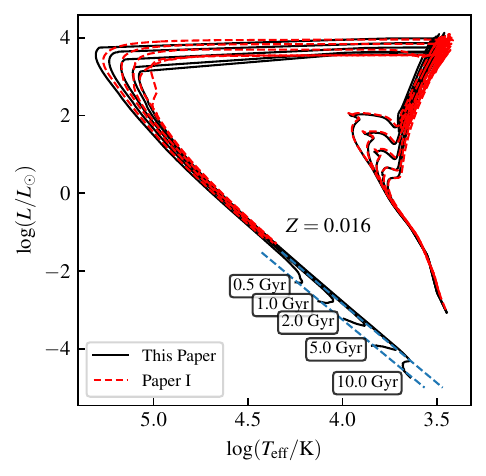}
  \caption{Isochrones at solar metallicity from MIST v1 and v2 (this
  work). The dashed blue tracks on the WDCS show the locations of
  $0.6\,M_\odot$ and $1.0\,M_\odot$ WD cooling tracks.}
  \label{fig:isochrones}
\end{figure}

\begin{figure*}
  \centering
  \includegraphics{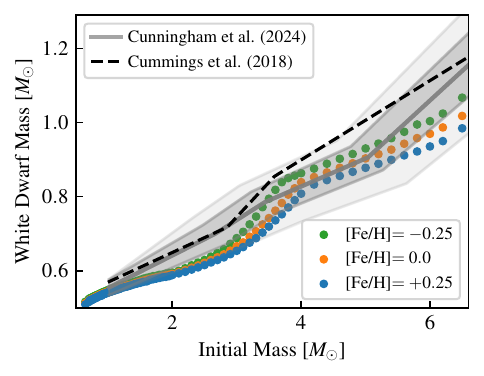}
  \includegraphics{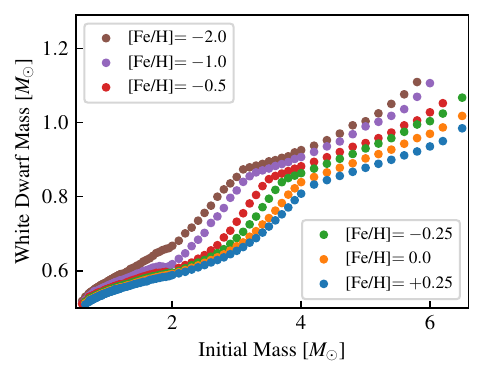}
  \caption{IFMR for MIST model grids over a range of metallicities,
    with $\afe = 0$ for all models shown here. The left panel includes
    the empirical IFMR inferences of \cite{Cummings2018} and
    \cite{Cunningham2024} for comparison to $-0.25 < \feh < +0.25$,
    including $\pm 1\sigma$ and $\pm 2\sigma$ error bars from
    \cite{Cunningham2024}. The right panel includes several
    lower-metallicity grids to show more of the full range of the IFMR
    from our model grids. Note that the green points
    for $\feh = -0.25$ here correspond to the grid of tracks plotted
    in the HRD and CMD in Figure~\ref{fig:HRtracks_all}.}
  \label{fig:IFMR}
\end{figure*}

We also include the WDCS in MIST isochrones for our model grids
with $\feh \geq -2.50$. As in Paper~II, very low-metallicity models do
not have a high success rate evolving
through the TP-AGB phase, so isochrones for $\feh < -2.5$ end
at the beginning of the TP-AGB, though the available model tracks
still extend to wherever the stellar evolution model reached. There
are some WD cooling model tracks even for $\feh < -2.5$, but the
isochrone files only include the WDCS for $\feh \geq 2.5$, where we
generally have $\sim$100 models that reached and evolved through this
stage at a given metallicity.
Figure~\ref{fig:isochrones} shows a collection of isochrones
including the WDCS for MIST v2 compared to v1. While Paper~I did
include the beginning of the WDCS, our tracks now extend to much older
ages and cooler models on the WDCS.

\subsection{Initial--Final Mass Relation} \label{s.IFMR}

Our model grids naturally produce an initial--final mass relation
(IFMR) in line with observational inferences (e.g.,
\citealt{Cummings2018,El-badry2018,Cunningham2024,Hollands2024}).
Figure~\ref{fig:IFMR} shows the IFMR produced by some of our model
grids over the metallicity range $-2 \leq \feh \leq +0.25$. The range
of WD masses seen in this figure appears to agree well with the
$\pm 1\sigma$ contours of the \cite{Cunningham2024} empirical IFMR,
which assumes a metallicity variation of $-0.25 < \feh < +0.25$.
On the low-mass end our models tend toward the lower end of the $\pm
1,2\sigma$ contours for metallicities near solar, and somewhat
lower-metallicity grids seem to agree better with the observational
inferences in that mass range.
  Figure~\ref{fig:IFMR} also shows the semi-empirical IFMR of
  \cite{Cummings2018} calculated using progenitor lifetimes from MIST
  Paper~I. This IFMR is somewhat systematically higher than our models
  here, but still generally agrees within its $1\sigma$ errors (not
  shown here to avoid confusion with the \cite{Cunningham2024} errors
  on the plot).

Generally, higher-metallicity models produce a ``shallower'' IFMR, meaning less
massive WDs for a given progenitor star mass. This is due to the
higher metallicity leading to opacities and stellar structure
configurations that drive stronger winds on the AGB, even though there
is no explicit metallicity dependence in the AGB wind prescription
\citep{Bloecker1995}. While the AGB wind prescription in our MIST
models may still have large uncertainties, the overall trend of a
shallower IFMR with higher metallicity is likely to be robust.
Other stellar evolution models have also indicated that
low-metallicity progenitor stars should produce more massive WDs
(e.g., \citealt{MillerBertolami2016,Lawlor2023}), while higher
metallicity leads to less massive WDs \citep{Byrne2025}.

\section{Stellar Evolution Physics} \label{s.physics}

We run all of our white dwarf models using MESA
\citep{Paxton2011,Paxton2013,Paxton2015,Paxton2018,Paxton2019,Jermyn2023}
version r24.08.1. The full MESA setup for running all of these WD
models is available at\dataset[10.5281/zenodo.15196934]{\doi{10.5281/zenodo.15196934}}.
This link includes directories containing starting models and MESA
work directories for each MESA WD model, organized according to
metallicity, with work directories named according to progenitor mass
in the MIST grid.

The MESA equation of state (EOS) is a blend of the OPAL \citep{Rogers2002}, SCVH
\citep{Saumon1995}, FreeEOS,\footnote{\url{https://freeeos.sourceforge.net/}}
HELM \citep{Timmes2000}, PC \citep{Potekhin2010}, and Skye
\citep{Jermyn2021} EOSes.
In particular, the Skye EOS covers the region of high density where
crystallization and C/O phase separation occur in WD interiors,
and includes the latent heat generated by crystallizing material \citep{Jermyn2021}.
Radiative opacities are primarily from OPAL \citep{Iglesias1993,
Iglesias1996}, with low-temperature data from \citet{Ferguson2005}
and the high-temperature, Compton-scattering dominated regime by
\citet{Poutanen2017}.  Electron conduction opacities are from
\citet{Cassisi2007} and \citet{Blouin2020apj}.
Nuclear reaction rates are from JINA REACLIB \citep{Cyburt2010}, NACRE \citep{Angulo1999} and
additional tabulated weak reaction rates \citet{Fuller1985, Oda1994,
Langanke2000}.  Screening is included via the prescription of \citet{Chugunov2007}.
Thermal neutrino loss rates are from \citet{Itoh1996}.

The following sections give more description of the particular
modeling choices made for MESA WD evolution in this work.

\subsection{Evolution over to the WD Cooling Sequence} \label{s.crossing}

The progenitor stellar evolution models from Paper~II end their
evolution when they leave the AGB and reach $T_{\rm eff} =
10{,}000\,\rm K$. At this point the models have a residual hydrogen
envelope of mass $\sim 10^{-3}\,M_\odot$. To enable the stellar
evolution model to contract and proceed across the HR diagram toward
the beginning of the WD cooling sequence, we start by taking these
models and temporarily applying an external pressure boundary
condition and turning off nuclear energy generation. We then allow the
models to evolve across the HR diagram toward hotter temperatures
until they reach $T_{\rm eff} > 60{,}000\,\rm K$.

Once the models have
reached this stage, we expedite the composition evolution by enhancing
the rate at which nuclear reactions occur by a factor of $10^3$ while
continuing to suppress the energy generation from nuclear
reactions. This allows the H and He envelopes to burn down to the
configurations that they will eventually have on the WDCS without
experiencing any burning instabilities. This suppresses any late or
very-late thermal pulses that might otherwise expel/consume the
residual hydrogen envelope, so all of our WD models evolve into WDs
with relatively ``thick'' hydrogen envelopes.
To verify that this procedure gives reasonable envelope
configurations, we also ran some model grids without these
modifications to the nuclear physics, and we found that for those
models that did not encounter instabilities, the final envelope
configurations and masses were very similar. For more on envelope
masses and configurations, see Section~\ref{s.composition}.
We turn on element diffusion during this phase to allow the high
surface gravity to begin the process of heavy element sedimentation
that will eventually form a pure hydrogen atmosphere.
We also allow thermohaline mixing \citep{Kippenhahn1980} during this
phase, which tends to flatten any molecular weight ($\mu$) inversion
(i.e., an oxygen abundance that increases outward in the core) that
might otherwise lead to later dynamical mixing in the core after the
WD cools further.

This phase of the evolution across the HR diagram continues until the
model has contracted to a radius of $0.1\,R_\odot$ (or $0.05\,R_\odot$
for $M_{\rm WD} > 0.7\,M_\odot$), at which point we begin the WD
cooling track. The model typically crosses to this stage in just a few
thousand years of evolutionary time or less, and starts the WD cooling track at
a luminosity of $L\sim 10^3-10^4\,L_\odot$ and temperature of
$T_{\rm eff} \gtrsim 100{,}000\,\rm K$ (see also
Figure~\ref{fig:HRtracks_all}).

\subsection{White Dwarf Cooling Evolution}

Once the model has reached a radius of $0.1\,R_\odot$ (or
$0.05\,R_\odot$ for models above $0.7\,M_\odot$), we begin the white
dwarf cooling track. We remove the extra surface pressure
so that the model can evolve with a physical surface boundary condition.
We continue to suppress energy from nuclear burning until the models
have further settled and reached a luminosity of $100\,L_\odot$,%
\footnote{This luminosity is chosen because it allows the envelope to
  settle into a structure that will be stable on the WDCS, while also
  being early enough that the timescale over which any suppressed
  burning would otherwise release energy should be very short. At most this
  burning could cause a delay of $M_{\rm H}Q_{\rm H}/Lm_p$, where
  $Q_{\rm H} \approx 7\,\rm MeV$ is the energy produced per H atom
  burned. For $M_{\rm H} \sim 10^{-3}\,M_\odot$ and $L > 100\,L_\odot$, this
  gives a delay of at most $10^6$~years, and in fact most of the
  burning typically happens at much higher luminosity, with
  substantially less than $10^{-3}\,M_\odot$ being consumed, so that
  the associated timescale is almost always shorter than
  $10^5$~years.}
at which point nuclear reactions are allowed to resume their normal
evolution. We evolve these models down the cooling track until they
reach $T_{\rm eff} = 2{,}000\,\rm K$, the edge of our tabulated
atmosphere boundary conditions.
For WDs that have convective surface layers, we employ the ML2 mixing
length theory prescription \citep{Bohm1971}, with
$\alpha_{\rm MLT} = 0.8$ to approximate the average results from the
3D simulations of \cite{Tremblay2015}.

We include the following pieces of physics in the models that affect the
evolution timescales on the WD cooling sequence, and we therefore give
a brief list along with references for the adopted default treatment
if relevant, followed by more discussion and references for each piece of physics:
\begin{itemize}
\item $^{22}$Ne Sedimentation \citep{Caplan2022}
\item Crystallization, C/O Phase Separation \citep{Bauer2023}
\item Electron Conduction Opacities \citep{Cassisi2007}
\item Surface Boundary Conditions \citep{Rohrmann2012}
\item Residual Hydrogen Burning (Section~\ref{s.envelopes})
\end{itemize}

{\it $^{22}$Ne Sedimentation}:
Much of the initial $\alpha$-element metallicity will end up as
$^{14}$N by the end of the main sequence due to CNO burning, which
will then be processed to $^{22}$Ne via $\alpha$ captures during He
burning. This $^{22}$Ne distributed through the WD core is slightly
heavier than surrounding elements due to the extra two neutrons in the
nucleus, and its sedimentation is an energy source that can delay cooling
\citep{Bildsten2001,Deloye2002,Garcia-Berro2008,Garcia-Berro2010,Garcia-Berro2011,Althaus2010,Bauer2020,Camisassa2021}.
Our MESA models include diffusion and sedimentation, and by default
include any heating associated with the sedimentation of
$^{22}$Ne or other heavy elements. The rate of sedimentation in dense WD interiors is governed
by the diffusion coefficients of \cite{Caplan2022}, which agree well
with the commonly adopted coefficients of \cite{Hughto2010}. These
coefficients yield relatively slow diffusion, so that the rate of
heating from sedimentation is low for all but the most metal-rich
compositions. This effect will therefore be negligible for most
of our models with $\feh < 0$, but can introduce small but noticeable
delays for $\feh \geq 0$, $\afe \geq 0$. Our treatment of heavy element
sedimentation also captures any heating associated with other heavy
elements that may be present, such as $^{26}$Mg or $^{56}$Fe, but
their effects are small compared to $^{22}$Ne.
Our models do not include any
diffusion in the solid state, which should be much slower than
diffusion in gas and liquid phases, though it may not be altogether
zero \citep{Caplan2024}. We also do not include any phase separation
or distillation of trace heavy elements such as $^{22}$Ne or $^{56}$Fe
\citep{Bauer2020,Caplan2020,Caplan2021,Blouin2021apjl,Caplan2023,Vanderburg2025}, which could lead to more
dramatic transport and associated cooling delays from these elements,
and may therefore explain more significant WD cooling anomalies such
as the $Q$-branch and NGC~6791 \citep{Cheng2019,Shen2023,Salaris2024,Bedard2024}.
These may also be addressable with MESA models as an extension of this
work, but are left for future exploration.

{\it Crystallization, C/O Phase Separation}:
White dwarfs eventually crystallize as their interiors cool
\citep{Salpeter1961}, and this releases latent heat
\citep{vanHorn1968,Tremblay2019}. The Skye EOS in MESA self-consistently calculates
the latent heat associated with this phase transition in WD interiors
\citep{Jermyn2021,Bauer2023}. This phase transition also leads to a
discontinuity of the C/O composition that excites further mixing in
the surrounding mantle and releases energy that can delay WD cooling
further
\citep{Stevenson1977,Mochkovitch1983,Isern1991,Segretain1993,Isern1997,Salaris1997,Montgomery1999}.
The MESA implementation of this process employs the phase diagram of
\cite{Blouin2020,Blouin2021}, and is described in detail in \cite{Bauer2023}.

{\it Electron Conduction Opacities}:
Heat transport is dominated by electron conduction in WD
interiors. In the deep interior, this conduction is very efficient, so
that the C/O core of a cooling WD often has a nearly isothermal
profile, and the models are not particularly sensitive to the details
of electron conduction there. However, the cooling timescales can be
sensitive to the details of electron conduction near the transition
between the degenerate core and the non-degenerate envelope
\citep{Blouin2020apj,Cassisi2021,Salaris2022,Salaris2024}. MESA
includes options to employ the conductive opacities of
\cite{Cassisi2007}, and to include additional corrections from
\cite{Blouin2020apj}. Significant uncertainties remain, and here we
adopt \cite{Cassisi2007} as the default conductive opacities for our
models. For more discussion on the state of the uncertainties, we
refer the reader to \cite{Cassisi2021} and \cite{Salaris2024}, and we
note that in principle it would be straightforward to produce updates
to our modeling for updated conductive opacity treatments as the need
arises.

{\it Surface Boundary Conditions}:
WDs have atmospheres that are not well-approximated by
gray radiative transport near the photosphere, and this can affect the
cooling timescales for cool WDs ($T_{\rm eff} \lesssim 6{,}000\,\rm K$).
\cite{Rohrmann2012} provides a set of tabulated surface boundary
conditions from calculations that include frequency-dependent
radiative transport, and we apply these boundary conditions for our WD
models.

\subsection{Bolometric Corrections} \label{s.BCs}

The MESA evolution produces theoretical surface properties specified in
terms of $\log g$, $T_{\rm eff}$, and $\log L$. We also provide
bolometric corrections for the WD regime using the DA/DC WD spectra of
\cite{Tremblay2011}, which are available at
\url{https://warwick.ac.uk/fac/sci/physics/research/astro/people/tremblay/modelgrids/}.
These synthetic spectra cover $1{,}500\,{\rm K} \leq T_{\rm eff} \leq
140{,}000\,\rm K$ and $6.5 \leq \log g \leq 9.5$, assuming pure
hydrogen composition.
  As confirmation that the atmosphere models used for our bolometric
  corrections are well calibrated, we note that the 0.6~$M_\odot$ WD
  model track at the peak of the WD mass distribution aligns very
  closely with the ``A branch'' peak density of WDs on the 2D
  histogram data shown in the lower-right panel of
  Figure~\ref{fig:HRtracks_all} \citep{GaiaHRD,Blouin2023}.
Our isochrones and model tracks are available
with synthetic photometry for all of the filters that are part of MIST.

\section{Composition} \label{s.composition}

One of the advantages of producing WD cooling models within the
framework of the MIST stellar evolution grids is that all WD models
have realistic composition profiles produced by stellar evolution
calculations of prior stages, and each of these models connects
directly back to its progenitor evolution.

\subsection{Core Composition}

\begin{figure}
  \includegraphics{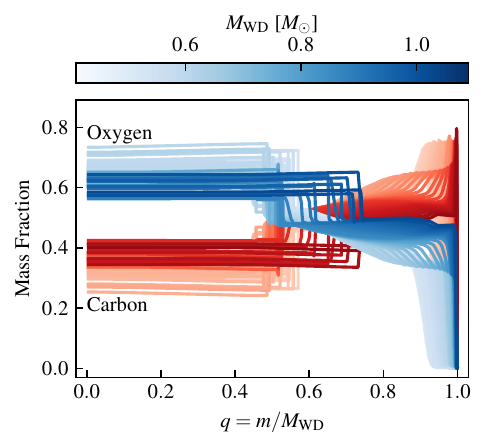}
  \caption{Carbon and oxygen mass fraction profiles for the same grid
    of WDs as shown in Figure~\ref{fig:HRtracks_all}. Darker lines
    represent more massive WDs from the model grid.}
  \label{fig:CO_cores}
\end{figure}

For C/O WDs, the primary components of the core composition are
$^{12}$C and $^{16}$O produced by prior core He burning, and $^{22}$Ne
produced from the progenitor CNO metallicity with a mass fraction roughly
equal to the progenitor metallicity. Our MIST progenitor models evolve
through all of the phases of burning that produce these compositions
in the WD interior, including the computationally challenging
thermally-pulsing AGB (TP-AGB) stage that builds the last 10--30\% of the C/O
core mass (Paper~I). Our models therefore naturally
contain interior $^{22}$Ne abundance reflecting the progenitor
metallicity in the grids.

Figure~\ref{fig:CO_cores} shows the C/O interior
profiles for our grid of models descended from progenitors of $\feh =
-0.25$, $\afe = 0$, the same models for which we plot the evolutionary
tracks in Figure~\ref{fig:HRtracks_all}.
Our WD models have central oxygen mass fractions in the range
$0.55-0.75$, with the trend being toward somewhat lower central oxygen
abundance as WD mass increases. This range of core C/O abundances
generally matches the state of the art for stellar evolution models,
though there are still substantial uncertainties associated with both
the $^{12}$C($\alpha$,$\gamma$)$^{16}$O reaction rate and the details
of core convection and mixing during the He burning phase
\citep{Kunz2002,Straniero2003,Fields2016,deBoer2017,Blouin2024}.
While asteroseismology has shown some promise in further constraining
C/O composition profiles
\citep{Metcalfe2001,Metcalfe2002,Metcalfe2003,Giammichele2018,Giammichele2022},
preliminary results sometimes produce central oxygen abundances that
are difficult to explain with current understanding of stellar
evolution physics \citep{DeGeronimo2017,DeGeronimo2019}, and it
appears more questions remain before seismology can definitively
shrink these uncertainties further
\citep{Fontaine2002,Timmes2018,Bell2019,Bell2022,Chidester2021,Chidester2022,Chidester2023}.
For now, we believe that the C/O abundance profiles produced by our
stellar evolution models provide a reasonable representation of our
current best estimate of WD compositions, and offer a significant
improvement over the more crude estimate of 50/50 (by mass) C/O core
composition that has been adopted for some widely used WD cooling
model grids in the past \citep{Fontaine2001,Bedard2020}.

\begin{figure}
  \includegraphics{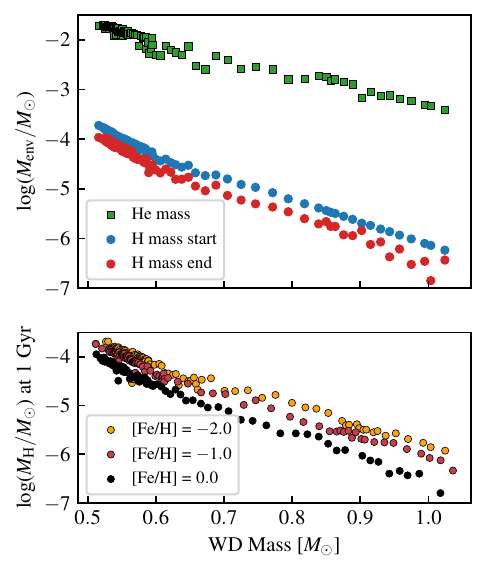}
  \caption{H and He envelope masses. The upper panel shows envelopes
    for the same $\feh = -0.25$ grid of models as shown in
    Figures~\ref{fig:HRtracks_all} and~\ref{fig:CO_cores}. For the H
    envelopes, the start mass is evaluated shortly after the model
    settles onto the beginning of the cooling track at a luminosity of
    $100\,L_\odot$, and the end mass is evaluated after the model has
    cooled to $2{,}000\,\rm K$ ($L < 10^{-5}\,L_\odot$).
      The lower panel shows the H envelope massses for grids at
      different metallicity $\feh$ ranging from $0.0$ to $-2.0$,
      evaluated at 1 Gyr of WD cooling age.
    }
  \label{fig:envelope_mass}
\end{figure}

\begin{figure}
  \includegraphics{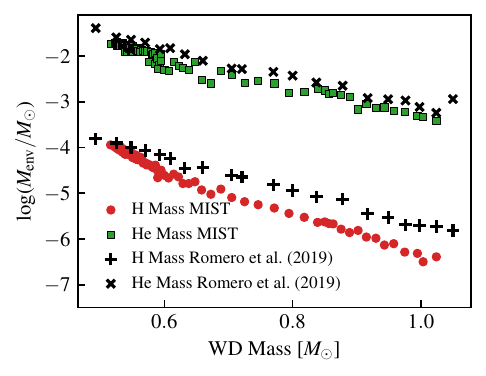}
  \caption{Comparison between envelope masses of MIST and
    \cite{Romero2019}. The \cite{Romero2019} masses are taken from
    their Table~1, which is given for WDs descended from $Z=0.01$
    progenitors that have cooled to $T_{\rm eff} = 12{,}000\,\rm K$.
    The MIST models here are taken from the same model grid as
    Figure~\ref{fig:envelope_mass}, but evaluated along a WD cooling
    age contour at 500~Myr, where they have similar temperature to the
    \cite{Romero2019} grid.}
  \label{fig:Romero_comparison}
\end{figure}

\subsection{Envelope Structure} \label{s.envelopes}

Another advantage of realistic progenitor evolution is that diffusion
and burning allow the envelopes to settle into more realistic
configurations than those of model grids with ad-hoc composition
profiles. Many WD model grids (e.g., \citealt{Bedard2020,Salaris2022})
adopt fixed H and He envelope masses of $10^{-4}M_{\rm WD}$ and
$10^{-2}M_{\rm WD}$ respectively. While these masses are a good
approximation based on evolutionary models that produce $0.6\,M_\odot$
WDs, these masses are too large for more massive WDs, which should
further burn away the bases of their envelopes before settling into a
stable configuration on the cooling track. The upper panel of
Figure~\ref{fig:envelope_mass} shows the H and He envelope masses
produced by our model grid descended from progenitors of $\feh =
-0.25$, $\afe = 0$, the same models for which we plot the evolutionary
tracks in Figure~\ref{fig:HRtracks_all}. The envelope masses are
specified in terms of the total mass of H and He in the model, and
the H mass at the start is evaluated when the WD reaches a luminosity of
$100\,L_\odot$ on the cooling track. While these models do have
envelopes on the order of $10^{-4}M_{\rm WD}$ (H) and $10^{-2}M_{\rm
  WD}$ (He) for masses around $0.6\,M_\odot$, more massive WDs have
envelopes that can be 1--2 orders of magnitude smaller.
  The lower panel of Figure~\ref{fig:envelope_mass} shows the
  dependence of the H envelope mass on progenitor metallicity.

Some DA WDs
are known to have much thinner envelopes from asteroseismology, but
stellar evolution cannot produce stable envelopes much more
massive than the ones shown here (e.g., \citealt{Romero2019}), so
these models should still be considered as having ``thick'' envelopes.
  We also provide Figure~\ref{fig:Romero_comparison} to compare
  our models to \cite{Romero2019}, which tabulates the maximum achievable
  H envelope masses for H atmosphere WDs at
  $T_{\rm eff} = 12{,}000\,\rm K$.
  To facilitate comparison to the data available from the
  \cite{Romero2019} models for this figure, the MIST models are taken
  from the model grid at $\feh =-0.25$ along a 500~Myr contour of
  constant WD cooling age, where they all have $T_{\rm eff}$ within
  a few thousand K of $12{,}000$. These models therefore come from
  the same model grid as the upper panel of
  Figure~\ref{fig:envelope_mass}, but are evaluated at different ages.

\begin{figure}
  \includegraphics{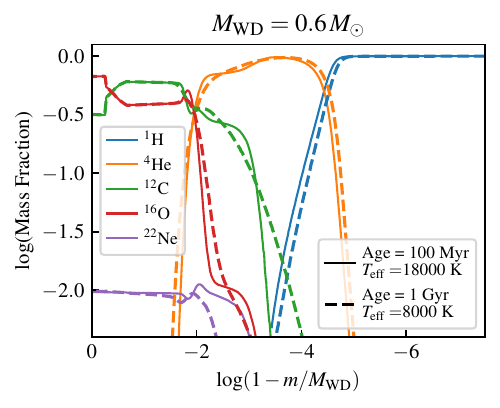}
  \includegraphics{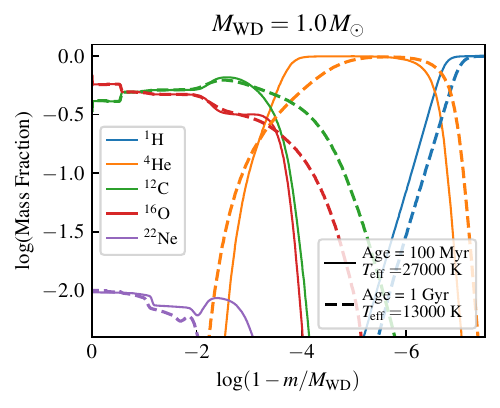}
  \caption{Composition profiles for $0.6\,M_\odot$ (upper) and
    $1.0\,M_\odot$ (lower) WD models at 100~Myr (solid) and 1~Gyr
    (dashed), emphasizing the H and He envelopes by plotting against
    the log of exterior mass coordinate.}
  \label{fig:envelope_profiles}
\end{figure}

Residual H burning at the base of the H envelope depletes the total H
mass over the course of the cooling
evolution. Figure~\ref{fig:envelope_profiles} shows some examples of
the composition profiles in the envelopes, and how they evolve with
time from 100~Myr to 1~Gyr thanks to both diffusion and nuclear
burning. Some of this H burning is enhanced by CNO reactions from the
overlap of the C and H diffusive tails in the He layer, the details of which are
governed by the history of time-dependent diffusion as C sinks away
from the surface, and later when the shape of both H and C diffusive tails
change as the deeper layers of the envelope become more degenerate at
cooler temperatures.

\begin{figure}
  \includegraphics[width=\apjcolwidth]{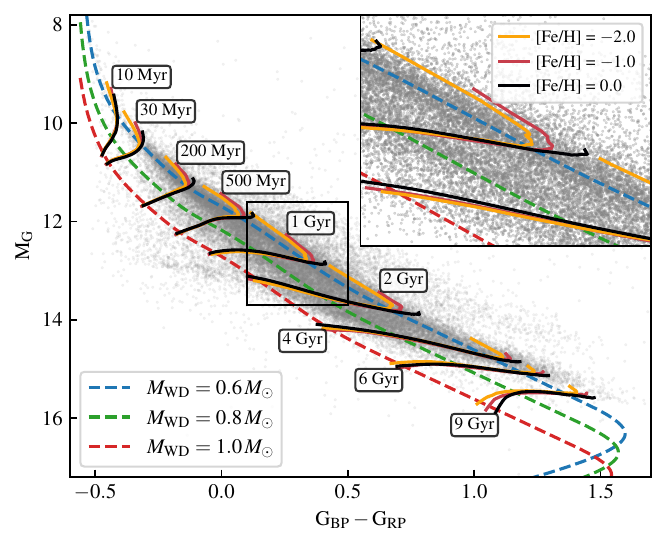}
  \caption{Example of delayed cooling for $M<0.6\,M_\odot$ WDs due to
    residual hydrogen burning in the envelope for WDs descended from
    low-metallicity ([Fe/H] = $-1,-2$) progenitors.}
  \label{fig:burndelay}
\end{figure}

For most models near solar metallicity,
this residual H burning does not have a large impact
on the cooling, contributing luminosities that are usually at most a
few percent of the radiated luminosity in the relatively early
evolution ($10^{-3}\,L_\odot < L < 10^{2}\,L_\odot$), and much less
later on. However, at lower metallicity ($\feh \lesssim -1$),
prior stellar evolution tends to leave a somewhat
thicker H envelope, and the residual burning of this envelope can
cause a significant cooling delay for $M_{\rm WD} < 0.6\,M_\odot$ WDs
that are less massive than average
\citep{Iben1986,MillerBertolami2013,Althaus2015}. Our model grids capture exactly
this effect, with somewhat more massive H envelopes for
lower-metallicity model grids (from less CNO burning, see
Figure~\ref{fig:envelope_mass}), and the cooling
age contours shown in Figure~\ref{fig:burndelay} demonstrate the
noticeable delay from residual H burning in WDs around $M_{\rm WD}
\approx 0.55\,M_\odot$ at $\feh = -1,-2$.

\section{Comparisons to Other Model Grids} \label{s.comparisons}

\begin{figure}
  \includegraphics{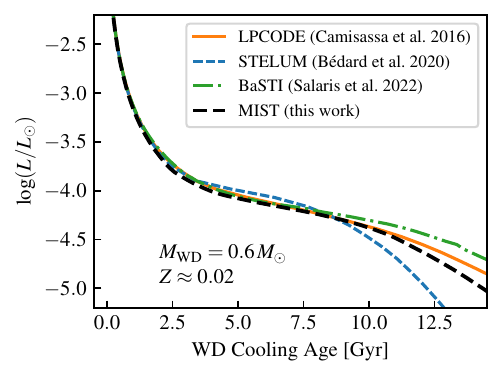}
  \includegraphics{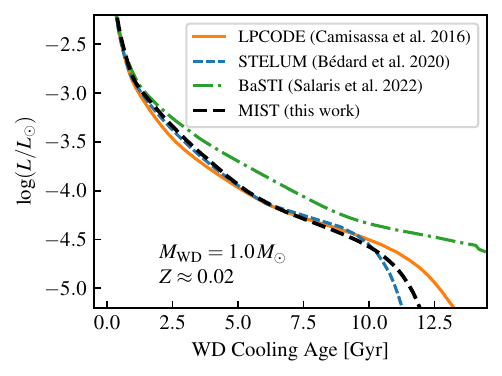}
  \caption{WD cooling timescales for model tracks of $0.6\,M_\odot$
    (upper panel) and $1.0\,M_\odot$ (lower panel) for models from
    LPCODE (orange), STELUM (blue), BaSTI (green), and this work
    (black).}
  \label{fig:compareCodes}
\end{figure}

Figure~\ref{fig:compareCodes} shows cooling timescales for
$0.6\,M_\odot$ and $1.0\,M_\odot$ WD models compared
with several other commonly used WD cooling model grids:
STELUM\footnote{\url{https://www.astro.umontreal.ca/~bergeron/CoolingModels/}}
\citep{Fontaine2001,Bedard2020,Bedard2022},
LPCODE\footnote{\url{https://evolgroup.fcaglp.unlp.edu.ar/modelos.html}}
\citep{Renedo2010,Althaus2015,Camisassa2016},
and BaSTI\footnote{\url{http://albione.oa-teramo.inaf.it/wdmod.php}}
\citep{Salaris2010,Salaris2022}.
For this comparison, we adopt models descended from approximately
solar metallicity ($Z \approx 0.02$) progenitors where applicable.
For our MIST models, this is the $\feh = \afe = 0$ model grid.
For LPCODE these are the $Z=0.02$ models of \cite{Camisassa2016}, and
for BaSTI these are the $Z=0.02$ models of \cite{Salaris2022}. For the
STELUM models of \cite{Bedard2020}, there is no sense of progenitor
metallicity, as the composition is set by hand and contains no trace
interior heavy elements such as $^{22}$Ne. For model grids that
have multiple conductive opacity options \citep{Salaris2022},
we have selected the models using the \cite{Cassisi2007} conductive opacities without the
\cite{Blouin2020apj} corrections, so that the conductive opacity
treatment is the same across all models being compared in this figure.

\begin{figure}
  \includegraphics[width=\apjcolwidth]{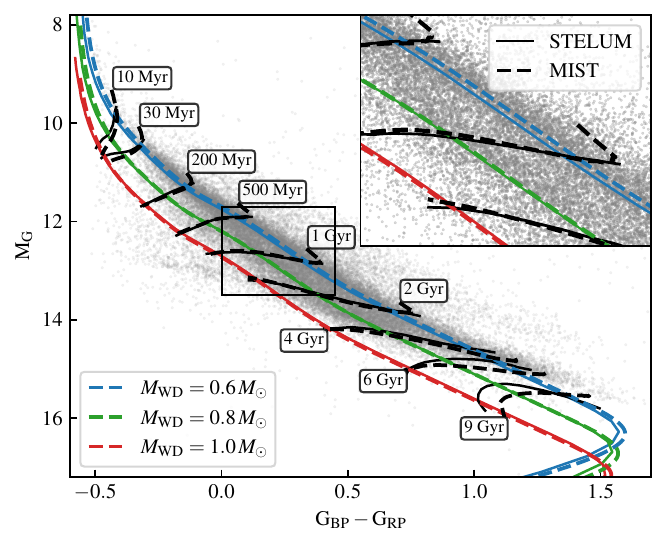}
  \caption{Comparison of model tracks and cooling age contours on the
    {\it Gaia} CMD for MIST (dashed) models and the STELUM
    (solid) models from \cite{Bedard2020}.}
  \label{fig:compareBedard}
\end{figure}

For a further detailed comparison against the STELUM models
specifically, Figure~\ref{fig:compareBedard} shows
cooling timescales and contours on the {\it Gaia} CMD for a
somewhat lower-metallicity MIST grid ($\feh = -0.5$, $\afe = 0$).
Cooling timescales agree very well up to about 2--4~Gyr,%
\footnote{The agreement is excellent except for $M_{\rm WD} \lesssim
  0.55\,M_\odot$ WDs where residual envelope burning alters the
  cooling timescale, see Section~\ref{s.envelopes}. The STELUM models
  do not include nuclear burning.}
after which our models tend to cool more quickly for
the next several Gyr.
  After $\approx 8-10$~Gyr, our models generally cool more slowly than
  STELUM models (as seen in Figure~\ref{fig:compareCodes}) regardless
  of metallicity.
We are able to very closely reproduce the \cite{Bedard2020} STELUM cooling timescales
with MESA models if we
  adopt the following assumptions that more closely match
  the modeling choices of the \cite{Bedard2020} model grid:
\begin{itemize}
\item Set composition by hand to be 50/50 C/O in the interior, with
  fractional envelope masses of $q({\rm H}) = 10^{-4}$, $q({\rm He}) =
  10^{-2}$.
\item Turn off all nuclear burning.
\item Use a gray Eddington-approximation $T(\tau)$ relation for the
  atmosphere boundary condition instead of the more detailed
  \cite{Rohrmann2012} boundary conditions.%
  \footnote{The \cite{Bedard2020} models do not use
  the Eddington $T(\tau)$ relation specifically, but they do employ a
  gray $T(\tau)$ relation, and we can closely reproduce the resulting
  cooling timescales with MESA using its Eddington-approximation
  atmosphere option.}
\item Alter the phase diagram so that crystallization occurs at
  $\Gamma \approx 175$, instead of the more realistic phase diagram from the
  Skye EOS \citep{Jermyn2021}.
\item Turn off C/O phase separation and the associated cooling delay
  \citep{Bauer2023}.
\end{itemize}
With the above changes, MESA model tracks agree very closely with
STELUM except for modestly faster evolution of the most massive WDs at
late times, which we speculate is likely due to the different EOS
for sufficiently cool WD interiors in the solid state at late times.
Since the above changes to the physics of our models all
represent simplified or more approximate treatments of WD cooling, we
believe our MESA models for MIST offer more physically realistic WD
cooling timescales.
  We also note that the above physical assumptions represent the choices
  made for the widely used \cite{Bedard2020} STELUM WD model grid, but
  they are not necessarily inherent to the STELUM code itself, which
  has also been employed with more sophisticated WD cooling physics
  options in more recent literature (e.g.,
  \citealt{Bedard2024,Vanderburg2025}).

\section{Conclusions}

We have introduced a major update to the MESA Isochrones and Stellar Tracks
library, extending the stellar evolution tracks down the WD cooling
sequence as DA/DC WDs with C/O cores in the mass range $\approx 0.5-1.05
M_\odot$. These models are now included in the MIST isochrones, and
both the model tracks and isochrones are publicly available on the
MIST project website. The WD models include state-of-the-art physics
for WD cooling timescales, including self-consistent composition
profiles produced by progenitor evolution, along with any related
cooling delays associated with residual nuclear burning, $^{22}$Ne
sedimentation, crystallization, and C/O phase separation.
This grid of models represents by far the largest available grid of
stellar evolution models from before the main sequence through the end
of the WD cooling sequence with detailed WD cooling physics.

Future extensions that could build upon this work may include
expanding the range of mass coverage for WD models up to higher masses
for both O/Ne and C/O WD models, and providing cooling sequences for
other WD spectral classifications such as DB WDs with He-dominated
atmospheres. As these grids of models are built on the publicly
available stellar evolution software MESA, they also provide a natural
starting point for investigations testing variations in the input
physics to WD cooling models. These MESA WD models could also be used as a
starting point for WD asteroseismology studies (e.g., \citealt{Timmes2018,Chidester2023})
with pulsation codes such as GYRE \citep{Townsend2013,Townsend2018}.

\vspace{1em}
{\it Acknowledgments:}
We thank Bill Paxton for developing and sharing MESA, and for
encouraging collaboration to extend its capabilities in the white
dwarf regime. We thank Antoine B{\'e}dard for helpful correspondence
clarifying the physical modeling choices made in STELUM models.
We thank Warren Brown for many insights and encouragements during the
development of this project. We also thank Saavik Ford and Julianne
Dalcanton for an enlightening discussion identifying the first
appearance of a white dwarf on the HR diagram.
This work was performed in part under the auspices of the
U.S. Department of Energy by Lawrence Livermore National Laboratory
under Contract DE-AC52-07NA27344.
This work benefited from discussions at the KITP program
{\it White Dwarfs as Probes of the Evolution of Planets, Stars, the
  Milky Way and the Expanding Universe} in the Fall of 2022,
and was supported in part by the National Science Foundation under
Grant No.\ NSF PHY-1748958.
TC was supported by NASA through the NASA Hubble Fellowship grant
HST-HF2-51527.001-A awarded by the Space Telescope Science Institute,
which is operated by the Association of Universities for Research in
Astronomy, Inc., for NASA, under contract NAS5-26555.

\software{%
  \texttt{MESA} \citep{Paxton2011, Paxton2013, Paxton2015, Paxton2018,
    Paxton2019, Jermyn2023},
  \texttt{Skye} \citep{Jermyn2021},
  \texttt{matplotlib} \citep{hunter_2007_aa},
  \texttt{SciPy} \citep{Scipy2020},
  \texttt{NumPy} \citep{der_walt_2011_aa},
  \texttt{statsmodels} \citep{statsmodels}, and
  \texttt{Python} from \href{https://www.python.org}{python.org}.
}

\end{document}